# Observation of huge thermal spin currents in magnetic multilayers

R. Ramos[1], T. Kikkawa[2,3], M. H. Aguirre[1,4,5], I. Lucas[1,6], A. Anadón[1,4], T. Oyake[7], K. Uchida[2,3,8], H. Adachi[3,9,10], J. Shiomi[7,8], P. A. Algarabel[4,11], L. Morellón[1,4], S. Maekawa[3,9,10], E. Saitoh[2,3,9,10,12],* and M. R. Ibarra[1,4,5],*

[1] Instituto de Nanociencia de Aragón, Universidad de Zaragoza, E-50018 Zaragoza, Spain.

[2]Institute for Materials Research, Tohoku University, Sendai 980-8577, Japan.

[3]ERATO, Japan Science and Technology Agency, Tokyo 102-0076, Japan

[4]Departamento de Física de la Materia Condensada, Universidad de Zaragoza, E-50009 Zaragoza, Spain.

[5]Laboratorio de Microscopías Avanzadas, Universidad de Zaragoza, 50018 Zaragoza, Spain.

[6]Fundación ARAID, 50018 Zaragoza, Spain.

[7]Department of Mechanical Engineering, The University of Tokyo, Tokyo 113-8656, Japan.

[8]PRESTO, Japan Science and Technology Agency, Saitama 332-0012, Japan.

[9]Advanced Science Research Center, Japan Atomic Energy Agency, Tokai 319-1195, Japan.

[10]CREST, Japan Science and Technology Agency, Sanbancho, Tokyo 102-0075, Japan.

[11]Instituto de Ciencia de Materiales de Aragón, Universidad de Zaragoza and Consejo Superior de Investigaciones Científicas, 50009 Zaragoza, Spain.

[12]WPI Advanced Institute for Materials Research, Tohoku University, Sendai 980-8577, Japan.

*Correspondence to: eizi@imr.tohoku.ac.jp or ibarra@unizar.es

**Thermal spin pumping constitutes a novel mechanism for generation of spin currents; however their weak intensity constitutes a major roadblock for its usefulness. We report a phenomenon that produces a huge spin current in the central region of a multilayer system, resulting in a giant spin Seebeck effect in a structure formed by repetition of ferromagnet/metal bilayers. The result is a consequence of the interconversion of magnon and electron spin currents at the multiple interfaces. This work opens the possibility to design thin film heterostructures that may boost the application of thermal spin currents in spintronics.**

Spin currents[1] are a central theme in spintronics[2, 3], they can be used to manipulate magnetic moment of nanomagnets[4], carry information[5] and produce magnetically controllable heat flow[6]. Therefore, generation and manipulation of spin currents are key aspects of spintronics. In this sense, the recently discovered spin Seebeck effect (SSE)[7] is of significant importance within this field, since it enables simple and versatile generation of spin currents from heat. The SSE consists on the thermal generation of a magnon spin current in a ferromagnetic material (F), which results in an injected electron spin current in an adjacent non-magnetic conductor (N) (Fig. 1a). This already known process is the result of a thermal non-equilibrium between magnons in F and conduction-electron spin accumulation in N. The spin current ($J_S$) is converted in the N layer into an electric field ($E_{ISHE}$) by means of the inverse spin Hall effect (ISHE)[8] as a result of the spin-orbit interaction. This has been explained theoretically using the linear-response approach[9, 10] and other methods[11-13].

The SSE has been experimentally established and observed in ferromagnetic materials: metals[7], semiconductors[14] and oxides[15]. This has contributed to the rapid expansion of the field of thermospin effects[16-18], which is focused on the study of the correlation between charge, heat and spin currents in magnetic materials. The observation of the SSE in magnetic insulators[15] is of great interest for potential applications as in energy harvesting[19], since it allows the generation of electric voltages without the thermal loss associated with mobile charge carriers. Another advantage of the SSE is that it involves the properties of at least two different materials that can be optimized independently. However, the SSE reported up to now is small[20] and improvement of the observed signal is still needed for application purposes. One approach is aimed at the enhancement of the spin current detection, by exploiting the spin-Hall angle characteristics of the N layer[3, 21]. Other possibility could be directed towards an increase of the generated spin current.

Here, based on the SSE, we investigate the thermal generation of spin current in magnetic multilayers formed by repeated growth of F/N bilayers, where F is magnetite ($Fe_3O_4$), an oxide extensively studied for spintronics[2, 3] due to its fascinating magnetic properties[22, 23], and N is platinum (Pt), a non-magnetic metal with strong spin-orbit coupling.

This type of system provides a unique template to investigate the physics of spin current at the interfaces, since it is expected to be enhanced in the multilayer samples (Fig. 1b), due to the boundary conditions for the propagation of the magnon/electron spin current through the F/N interfaces.

**Results**

We have systematically studied multilayers based on the repetition of $Fe_3O_4$/Pt bilayer structures grown on MgO (001) substrates (Fig. 2a) (see Methods for details).

We performed SSE measurements of the multilayer thin film samples as in the configuration described elsewhere (Supplementary Fig. 1 and Note 1); an external magnetic field $H$ is applied in the $x$ direction while a temperature gradient $\nabla T$ is sustained in the z direction by stabilizing a temperature difference $\Delta T$ between the bottom and top surfaces of the sample (see Fig. 1b). The voltage generated by the thermally induced $E_{ISHE}$ is measured by contacting the Pt top layer along the $y$ direction. It has been previously reported that despite $Fe_3O_4$ being electrically conductive at room temperature, the response observed in the $Fe_3O_4$/Pt system is highly dominated by the SSE[24]. This is due to the resistance mismatch between Pt and $Fe_3O_4$ layers, with $Fe_3O_4$ being two orders of magnitude more resistive than Pt. We have measured the contribution of the anomalous Nernst effect (ANE) of the magnetite layer, which accounts for less than 1 % of the measured voltage; this implies that the observed effect is dominated by the SSE of the $Fe_3O_4$/Pt bilayer, in good agreement with our previous study[24]. This is also supported by measurements varying the directions of applied magnetic field and thermal gradient with respect to the $Fe_3O_4$/Pt interface[25], which show the absence of both; ANE in $Fe_3O_4$/Pt bilayer and ANE due to transport magnetic proximity effect in Pt (Supplementary Fig. 2 and Note 2).

Furthermore, even if there was transport magnetic proximity effect in Pt, the voltage signal from such proximity effect would not scale with the number of layers because each layer is connected in parallel electrically. The upper limit of the transport magnetic proximity effect in the present multilayer system, which is inferred from the systematic study of the effect in a Pt/YIG system[25] multiplied by the number of multilayers, is far below the observed SSE enhancement.

Figure 2b shows the $H$ dependence at room temperature of the observed SSE voltage ($V_{ML}$) in the multilayer system normalized with the temperature difference ($\Delta T$) for different number of $Fe_3O_4$/Pt bilayers ($n$), showing that the signal scales up as $n$ is increased. The observed enhancement cannot be explained by the dependence of the SSE on the thickness of ferromagnetic material, as evidenced from the SSE voltage measured for a $Fe_3O_4$ (278 nm)/Pt(14 nm) sample, which is much lower than the value of the multilayer with $n = 6$ (Supplementary Fig. 3 and Note 3). This shows that the increased SSE is a consequence of the enhancement of the spin current at the interfaces of the multilayered structure. One scenario that should be discarded is the possible reduction of the thermal conductivity with the increasing number of layers as a consequence of the increased phonon scattering across multiple interfaces[26]. Thermoreflectance measurements showed a nearly constant value for the thermal conductivity upon increasing the number of bilayers. The average measured value was 2.9 ± 0.2 W/m-K, independent of the number of bilayers, thus the reduction of the thermal conductivity in the multilayer system is not the origin of our observations (Supplementary Fig. 6 and Note 4).

In order to understand the origin of the observed phenomenon, we investigate how the SSE is influenced by varying the nature and thickness of the non-magnetic interlayers. We performed measurements on a multilayer system in which the inner Pt interlayers of the samples were replaced by MgO, leaving only the topmost Pt layer. This structure does not disrupt the thermal conduction across the multilayer structure, due to the high thermal conductivity of MgO. Therefore, the heat transport is maintained while the electron/spin transport across the MgO electrical insulator spacer is suppressed. Figure 3a shows the comparative results obtained for 1 × ($Fe_3O_4$/Pt), 3 × ($Fe_3O_4$/Pt) and 2 × ($Fe_3O_4$/MgO)/$Fe_3O_4$/Pt sample systems at 300 K. It is interesting to observe that, despite having similar thermal conductivities, the multilayer system with MgO interlayers shows a strong reduction of the SSE signal as a consequence of no spin current propagation through the MgO interlayers. The obtained SSE value is comparable to the one obtained for a single $Fe_3O_4$/Pt bilayer. This result is a strong indication that the observed giant SSE is a purely spin current effect, in which the existence of multiple $Fe_3O_4$/Pt interfaces is a relevant factor.

To further study the role of the Pt interlayers in the enhancement of the spin current, we performed measurements of the SSE of a 6 × ($Fe_3O_4$/Pt) multilayer, with Pt layers of a thickness of 7 nm, which is of the order of the reported spin diffusion length of Pt[27]. This sample shows a giant SSE (~ 25 μV/K), with a huge increase respect to the one observed for a single bilayer sample (Fig. 3b). This result clearly indicates the relevance of the spin current propagation across the Pt interlayers due to the characteristic spin diffusion length of the metal.

**Discussion**

We will now present a model to qualitatively explain the physics behind the presented results. This model considers a magnon spin current in the F layer[5, 13]. We impose as boundary conditions: a) the spin/magnon current should vanish at the top/bottom surface of the multilayer structure, b) the continuity of the magnon/electron spin currents at the $Fe_3O_4$/Pt interfaces[28]. Under these conditions, an enhancement of the spin current in the multilayer system is predicted, with maximum value at the central interlayers of the structure (Supplementary Figs. 8 and 9 and Note 5). To explain our result, we need to consider the fact that the lateral dimensions of the samples are about six orders of magnitude larger than the thickness of the F/N bilayers, therefore the out-of-plane electrical resistance of F is negligibly small compared to the in-plane resistance, this is also supported by resistivity measurements in the multilayers and SSE measurements performed on F/N/F structures (Supplementary Figs. 10 and 11 and Notes 6 and 7). This implies that even if we are placing the electrical contacts in the topmost Pt layer, we are in fact contacting all the Pt layers simultaneously, thus observing an averaged SSE voltage. Figure 4 shows the comparison of the experimental values of SSE and the model predictions for the average spin current ($<J_S>$), we can clearly see that the model reproduces the main features of the observed SSE in F/N multilayers, demonstrating the enhancement of the spin current and consequently the spin Seebeck voltage upon increasing the number of multilayers.

The observed huge enhancement of the spin current deep inside the multilayer structure is a new phenomenon, consequence of the existence of multiple $Fe_3O_4$/Pt interfaces. This originates an increase of one order of magnitude on the SSE, raising the voltage to a value never reported in thin film based structures. These results pave the way to harness heat for its utilization as a source of spin currents in future generation of spintronic devices.

**Methods**

**Sample preparation**

The $Fe_3O_4$ films were grown by pulsed laser deposition (PLD) using a KrF excimer laser with 248 nm wavelength and 10 Hz repetition rate. The Pt films were deposited in the same UHV chamber by DC magnetron sputtering. The samples were characterized structurally, magnetically and electrically by XRD, TEM, SQUID and four probe resistivity measurements. Film cross section were prepared by Focused Ion Beam and measured by high angle annular dark field (HAADF) scanning transmission electron microscopy (STEM) image carried out in a probe aberration corrected FEI Titan 60-300 operated at 300 kV. Multilayer $n \times (Fe_3O_4/Pt)$ samples present a very well defined contrast, atomically sharp and with no inter-diffusion.

**Spin Seebeck effect (SSE) measurement setup**

The measurements were performed independently on the same samples: in a thermoelectric measurements probe compatible with an Oxford spectrostat NMR40 continuous flow He cryostat (Zaragoza) and in a physical property measurement system PPMS of Quantum design (Sendai), obtaining consistent results in both systems. The sample is placed between two AlN plates; a resistive heater is connected to the upper plate and the lower plate provides the heat sink and is in direct contact with the thermal link of the cryostat. A temperature gradient is generated by application of an electric current to the heater and the temperature difference between upper and lower plate is monitored by two T-type (E-type in PPMS) thermocouples. The AlN is used as an electrical insulating and thermal conducting material, to avoid electrical shorts in the sample. The samples were contacted with thin Al (Au in PPMS) wires with a diameter of 25 μm. To minimize thermal losses, the temperature of the wires is stabilized by thermal anchoring to the sample holder. The thermoelectric voltage was monitored with a Keithley 2182A nanovoltmeter. Sample dimensions for SSE measurements are: $L_x$ = 2 mm, $L_y$ = 7 mm and $L_z$ = 0.5 mm.

**Acknowledgments**

The microscopy works have been conducted in the "Laboratorio de Microccopías Avanzadas" at "Instituto de Nanociencia de Aragón – Universidad de Zaragoza". This work was supported by the Spanish Ministry of Science (through projects PRI-PIBJP-2011-0794 and MAT2011-27553-C02, including FEDER funding), the Aragón Regional Government (project E26) and Thermo-Spintronic Marie Curie CIG (Grant Agreement 304043). This research was also supported by Strategic International Cooperative Program, PRESTO "Phase Interfaces for Highly Efficient Energy Utilization", CREST "Creation of Nanosystems with Novel Functions through Process Integration", ERATO "Spin Quantum Rectification" from JST, Japan, Grant-in-Aid for Scientific Research on Innovative Areas "Nano Spin Conversion Science" (26103005), Grant-in-Aid for Scientific Research (A) (24244051), Grant-in-Aid for Young Scientists (A) (25707029), Grant-in-Aid for Challenging Exploratory Research (26600067) from MEXT, Japan, NEC Corporation, the Casio Science Promotion Foundation, and the Iwatani Naoji Foundation.

**Author Contributions**

R. R., T. K. and A. A. performed the measurements. R. R. and T. K. analysed the data; H. A. and S. M. provided the theoretical analysis; I. L and A. A contributed to the sample fabrication and characterization; K. U. and T. K. designed and implemented the SSE setup in Sendai. R. R. designed and implemented the SSE setup in Zaragoza; R. R., M. H. A. and A. A. contributed to the optimization of the SSE setup in Zaragoza; M. H. A. performed STEM-HAADF measurements and image analysis. T. O. and J. S. contributed to the thermal conductivity measurements. R. R., K. U., H. A. and J. S. wrote the manuscript; all authors discussed the results and commented on the manuscript; M. R. I., E. S. and S. M. planned and supervised the study.

**Additional Information**

**Supplementary Information** accompanies this paper at

http://www.nature.com/naturecommunications

**Competing financial interests**: The authors declare no competing financial interests.

**Figure 1 | Concept of the enhancement of thermal spin currents in magnetic multilayers.**
**(a)** Schematic illustration of the SSE in an individual F/N bilayer (where F and N are a ferromagnetic material and a non-magnetic metal, respectively). When a temperature gradient $\nabla T$ is applied to a magnetic film a magnon spin current is generated in the magnet. **(b)** Schematic illustration of the giant SSE in a magnetic multilayer formed by repetition of F/N bilayers. The thermally generated spin current is amplified in the central layers as a consequence of the continuity of the magnon and electron spin currents at the interfaces (the size of the arrows represents the magnitude of the spin current). The theoretically predicted dependence of the spin current ($J_s$) across the multilayer is sketched (the exact calculated dependence is outlined in supplementary text, section E). The giant SSE is detected due to the ISHE of the N layers.

**Figure 2 | Morphology and spin Seebeck results in Fe$_3$O$_4$/Pt multilayer** **(a)** High angle annular dark field (HAADF) scanning transmission electron microscopy (STEM) image of the cross section of a 6 × (Fe$_3$O$_4$($t_F$)/ Pt($t_N$)) sample, where $t_F$ = 34 ± 2 nm and $t_N$ = 17 ± 2 nm. Details of the Pt layer and MgO/Fe$_3$O$_4$ interface are shown to demonstrate the epitaxial growth of the multilayers. **(b)** $H$ dependence of the SSE in $n$ × (Fe$_3$O$_4$(34 nm)/Pt(17 nm)) measured at room temperature.

**Figure 3 | Influence of the nature of non-magnetic interlayers on the SSE in multilayers. (a)** Effect of Pt replacement by MgO (electrical insulator and thermal conductor) interlayers, the data shows the comparison of the SSE for 1 × (Fe$_3$O$_4$/ Pt), 3 × (Fe$_3$O$_4$/Pt) and 2 × (Fe$_3$O$_4$/MgO)/Fe$_3$O$_4$/Pt. The thicknesses of the layers are $t_{Fe3O4}$ = 34 ± 2 nm, $t_{Pt}$ = 17 ± 2 nm and $t_{MgO}$ = 8 ± 1 nm. **(b)** SSE measured in a Fe$_3$O$_4$($t_F$)/Pt($t_N$) bilayer compared with the giant SSE (GSSE) observed in a 6 × (Fe$_3$O$_4$($t_F$)/Pt($t_N$)) multilayer, where $t_F$ = 23 ± 1 nm and $t_N$ = 7 ± 1 nm.

**Figure 4 | Theoretical fit for the giant SSE in magnetic multilayers.** Comparison of the $n$ dependence between the measured SSE coefficient ($S_{SSE} = (V_{ML}^{sat}/\Delta T)(L_z/L_y)$) and the model prediction of the average spin current ($<J_S>$) for a multilayer structure with $n \times$ (Fe$_3$O$_4$(34 nm)/Pt(17 nm)), the spin current strength is normalized by - $L_m \nabla_z T$ (supplementary text, section E).

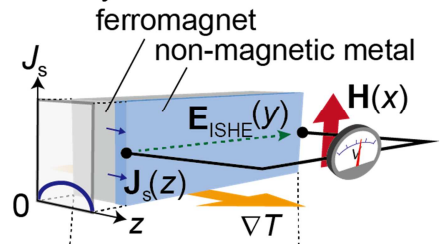
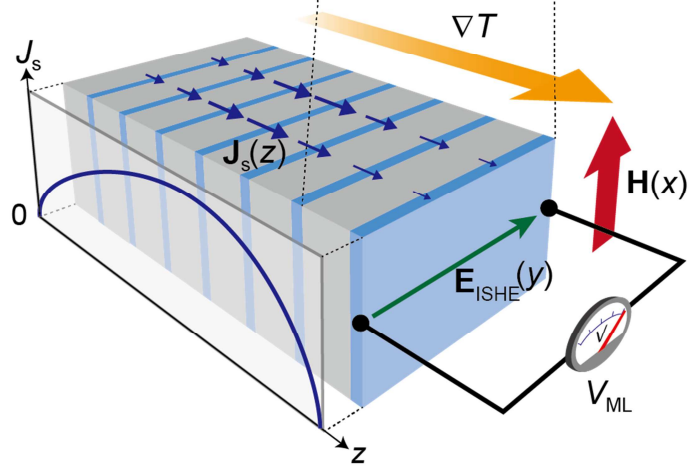

Fig. 1

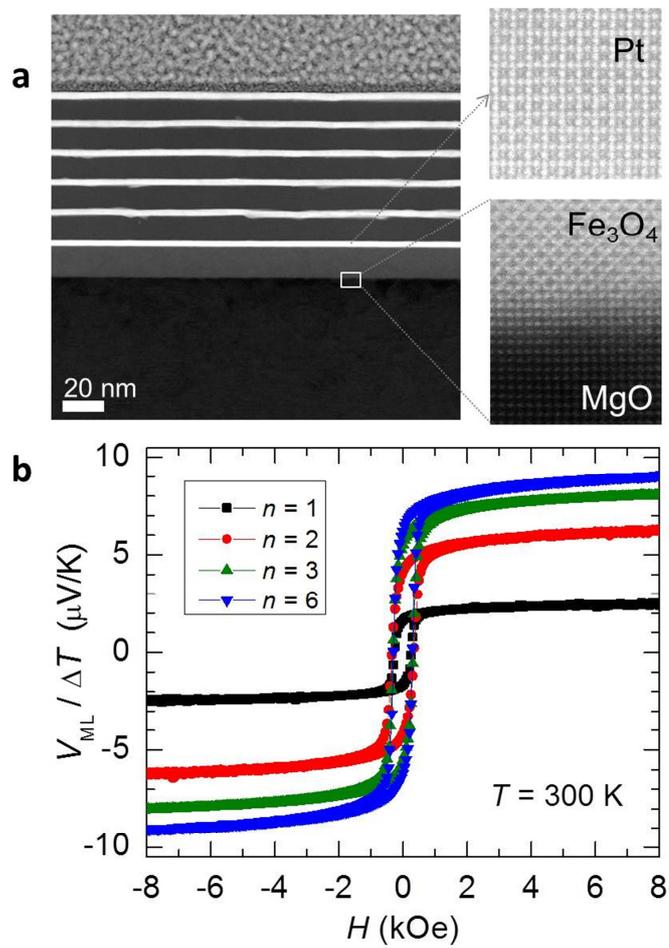

Fig. 2

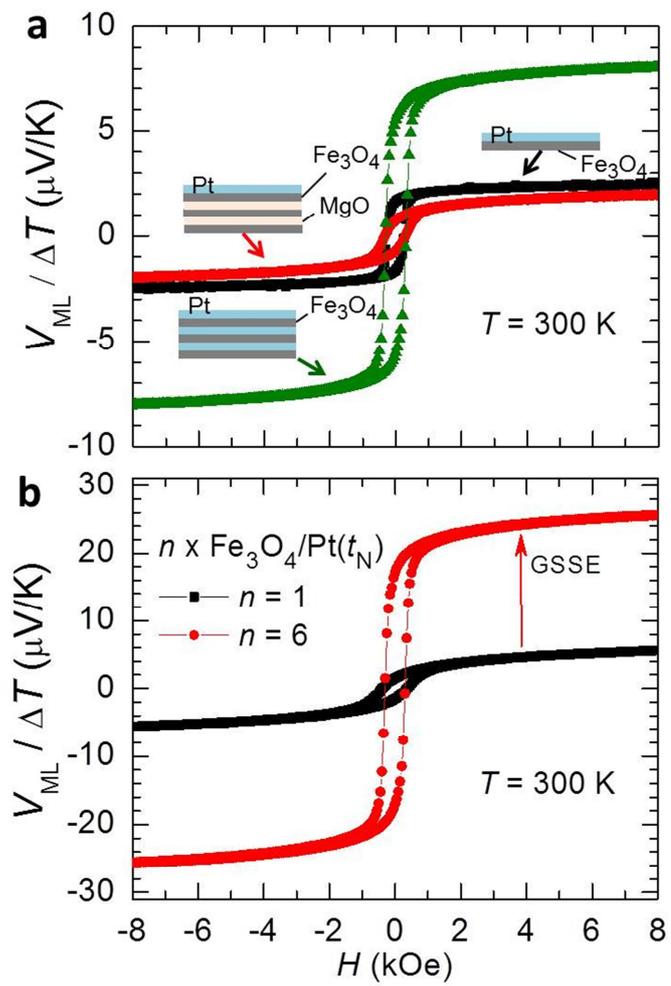

Fig. 3

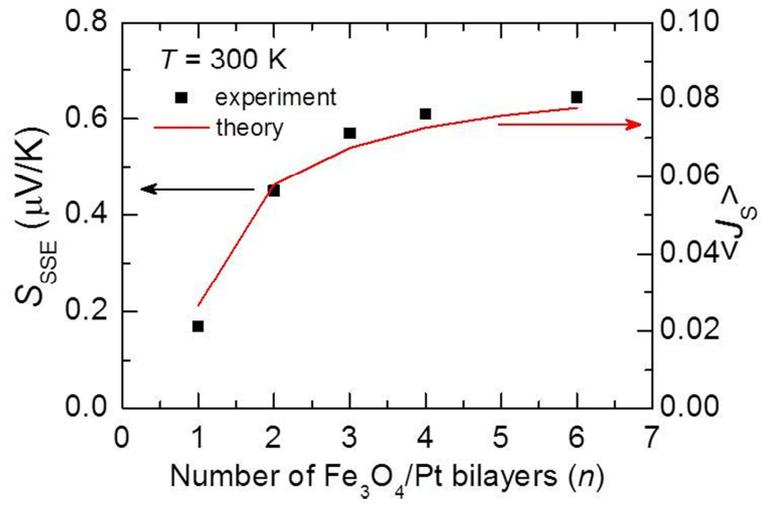

Fig. 4

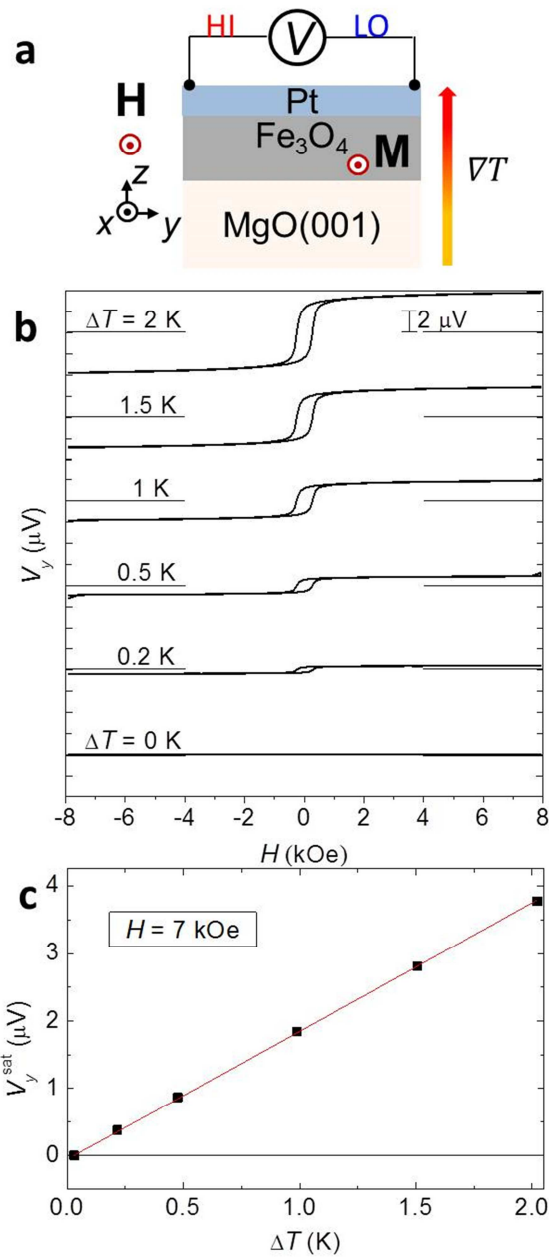

**Supplementary Figure 1 | Spin Seebeck effect of MgO(001)//Fe$_3$O$_4$/Pt. (a)** A schematic illustration of the measurement geometry for the longitudinal spin Seebeck effect. **(b)** Results obtained for different applied temperature differences ($\Delta T$) between top and bottom of the sample, measured at 300 K. **(c)** Dependence of $V_y^{\text{sat}}$ measured at 7 kOe for the different magnitudes of $\Delta T$ applied across the sample (the line shows the best fit to a linear dependence).

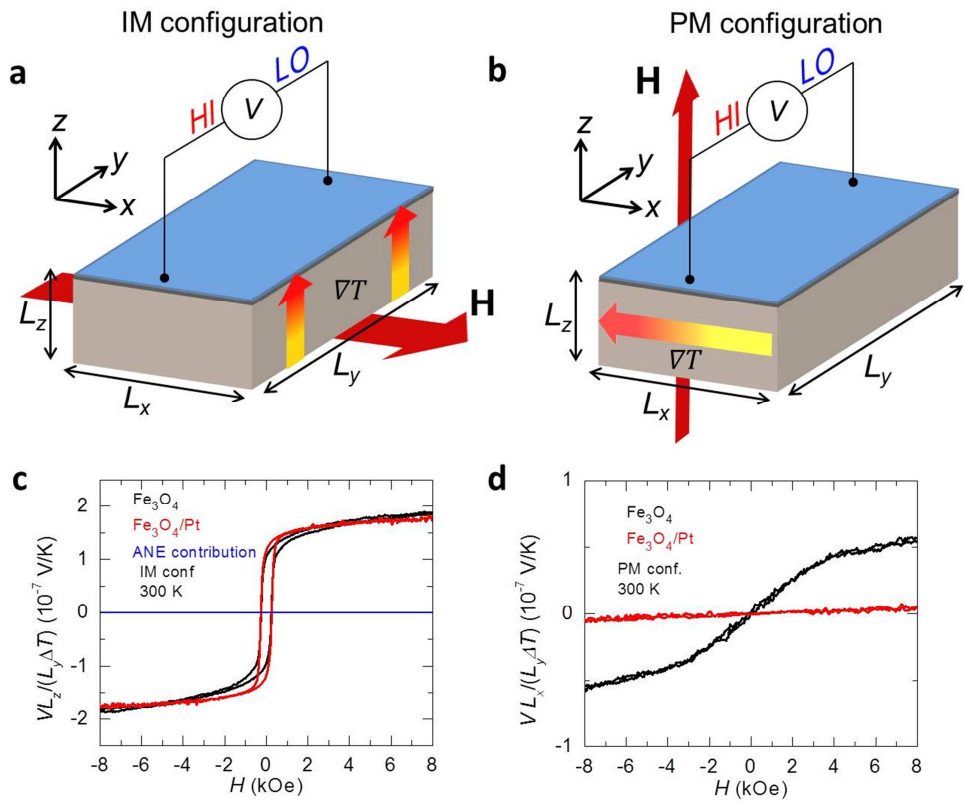

**Supplementary Figure 2 | Dependence of ANE and SSE on the direction of applied thermal gradient and magnetic field. (a)** Schematic illustration of the in-plane magnetized (IM) configuration where the SSE is present. **(b)** Schematic illustration of the perpendicular magnetized (PM) configuration where the SSE vanishes due to measurement geometry. **(c)** Results of ANE and SSE obtained from the measurement of a MgO(001)//Fe$_3$O$_4$(34) and MgO(001)//Fe$_3$O$_4$(34)/Pt(17) sample respectively, in the IM configuration (thicknesses in nm). The blue line shows the contribution of the ANE to the SSE measured signal using equation (S3). **(d)** Results obtained in the PM configuration. The ANE signal is still present; while the SSE vanishes due to measurement geometry (thermal spin pumping across the Fe$_3$O$_4$/Pt interface is parallel to the applied magnetic field).

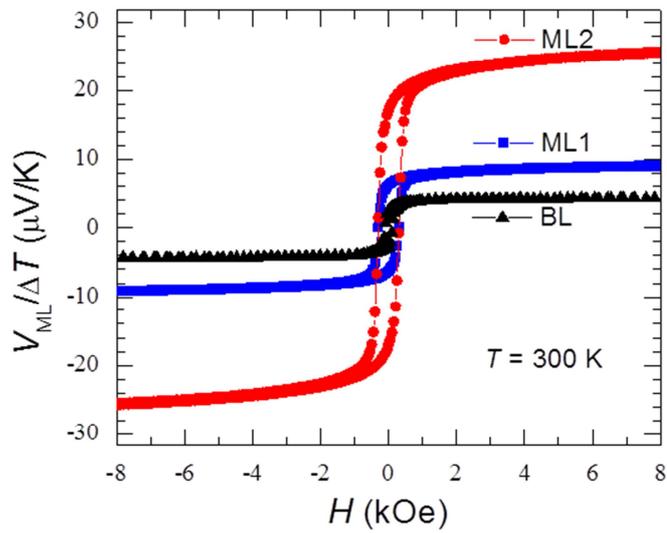

**Supplementary Figure 3 | Comparison of SSE between multilayers and a single bilayer with thick Fe$_3$O$_4$ film.** Observed SSE voltages for Fe$_3$O$_4$ (278 nm)/Pt (14 nm), 6 × Fe$_3$O$_4$(34 nm)/Pt(17 nm) and 6 × Fe$_3$O$_4$(23 nm)/Pt(7 nm) samples, labeled in the figure as BL, ML1 and ML2 respectively.

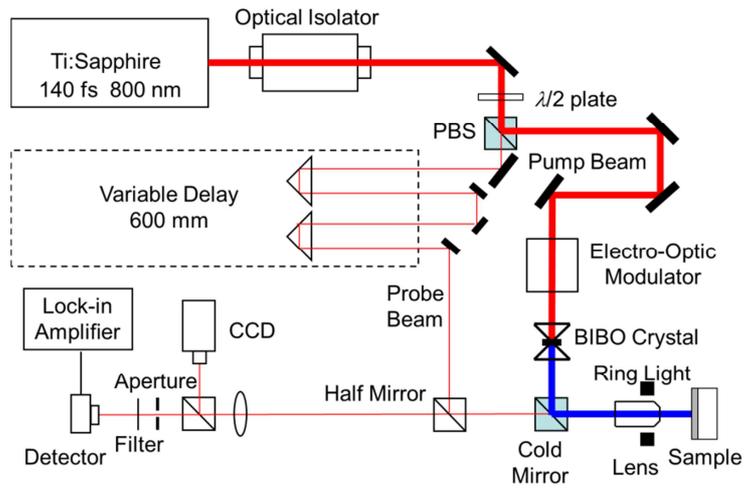

**Supplementary Figure 4 | A schematic diagram of the time domain thermoreflectance (TDTR) method.**

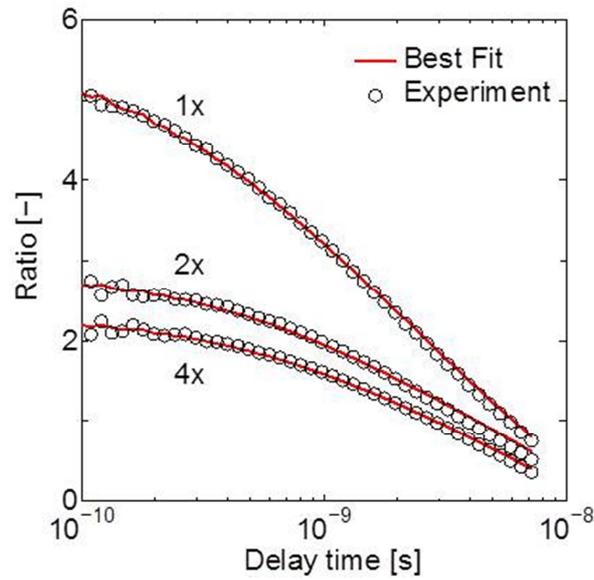

**Supplementary Figure 5 | Time histories of TDTR signals for $Fe_3O_4$ (34 nm)/Pt (17nm) multilayers with 1, 2, and 4 periods**. Circles and lines denote the experimental data and the least-square-fitted solutions of the analytical heat conduction model, respectively. The fitting parameters are the interfacial thermal conductance of Al/multilayer and thermal conductivity of multilayer. The moderate fluctuations in the experimental data at time less than 200 ps are due to acoustic echoes in the Al layer (*37*), which has been confirmed not to affect the fitting result.

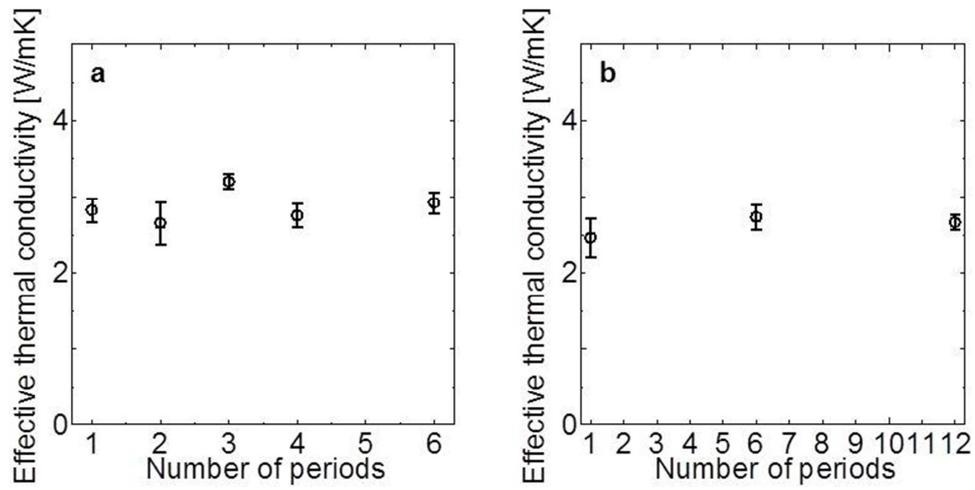

**Supplementary Figure 6 | Dependence of effective thermal conductivity of the multilayers on the number of $Fe_3O_4$/Pt period** for cases of **(a)** $Fe_3O_4$ (34 nm)/Pt (17 nm) and **(b)** $Fe_3O_4$ (23 nm)/Pt (7 nm) samples.

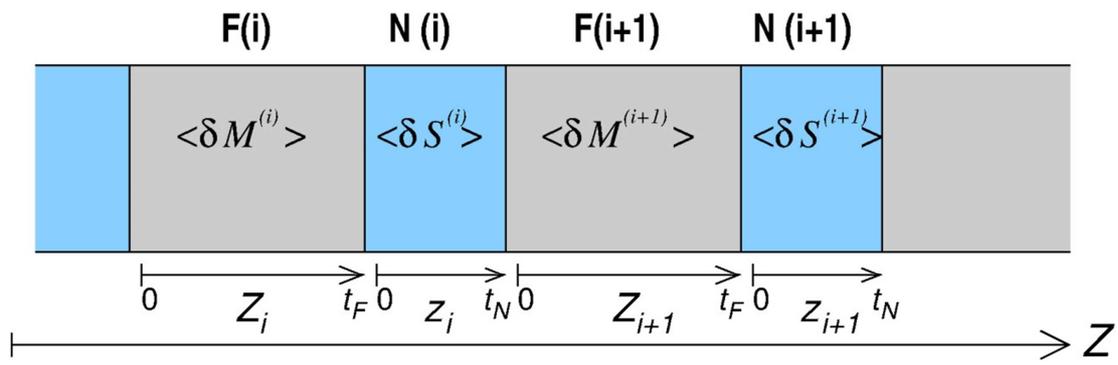

**Supplementary Figure 7 | The coordinate.** Schematic view of the multilayer system and its coordinate.

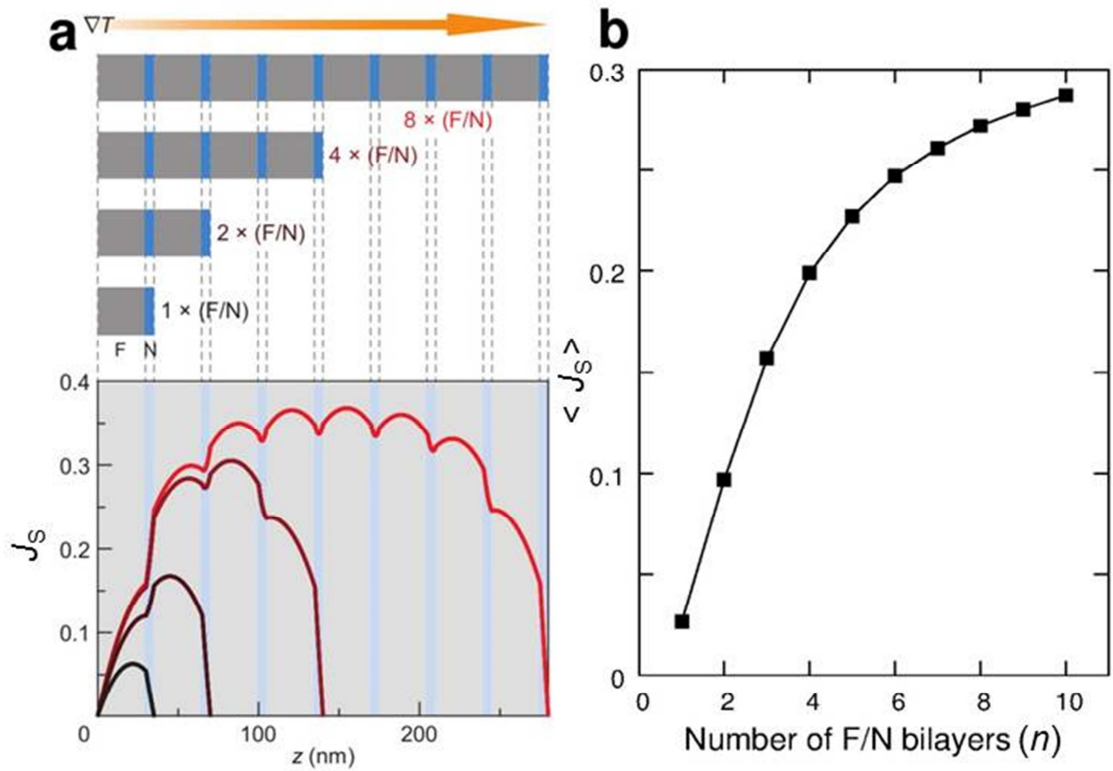

**Supplementary Figure 8 | Influence of multilayer structure on SSE. (a)** An example of the spin current profile calculated for a F(30 nm)/N(5 nm) multilayer system for several values of the number of bilayers, *n*. In F region, magnon spin current $J_m$ is also plotted. **(b)** Averaged spin current $\langle J_s \rangle$ as a function of *n*. The strength of the spin current is normalized by $-L_m \nabla_z T$.

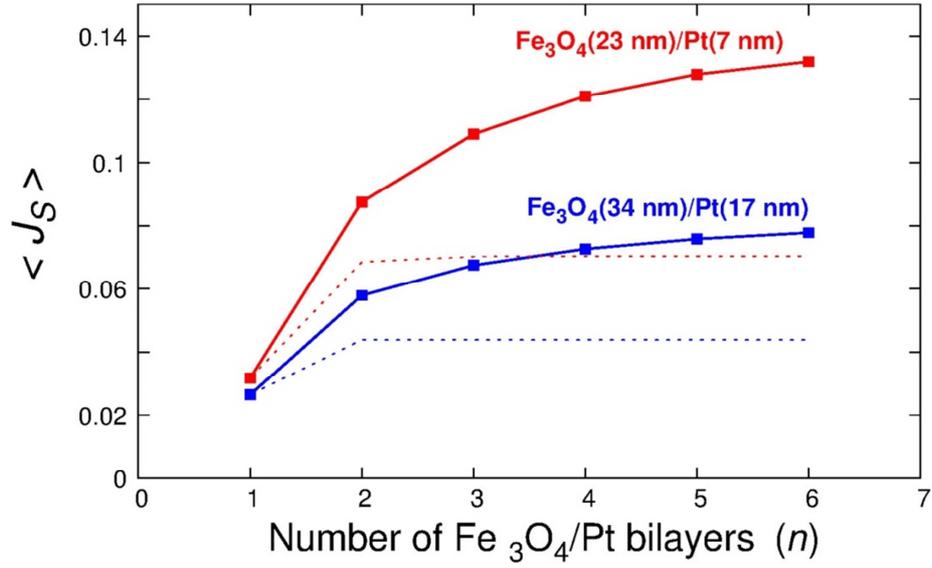

**Supplementary Figure 9 | Model calculation of the giant SSE in $Fe_3O_4$/Pt multilayers.** Averaged spin currents as a function of the number of bilayers in a $Fe_3O_4$(23 nm)/Pt(7 nm) system (red) and a $Fe_3O_4$(34 nm)/Pt(17 nm) system (blue) are shown. The dotted line represents the spin current in the toplayer only, and the spin current strength is normalized by $-L_m \nabla_z T$.

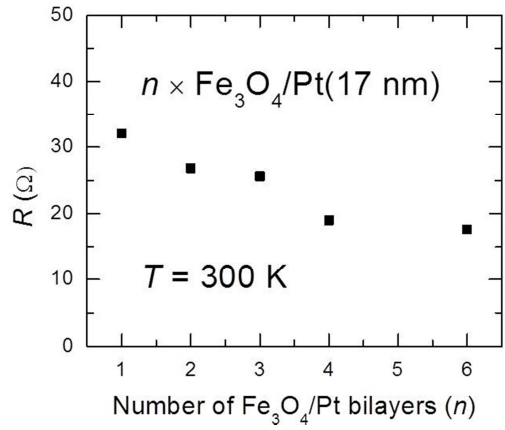

**Supplementary Figure 10 | Resistance vs number of Fe₃O₄/Pt bilayers (*n*)**. Resistance per 7 mm (sample length) for multilayers with $t_N$ = 17 nm. The applied current excitation is 10 μA.

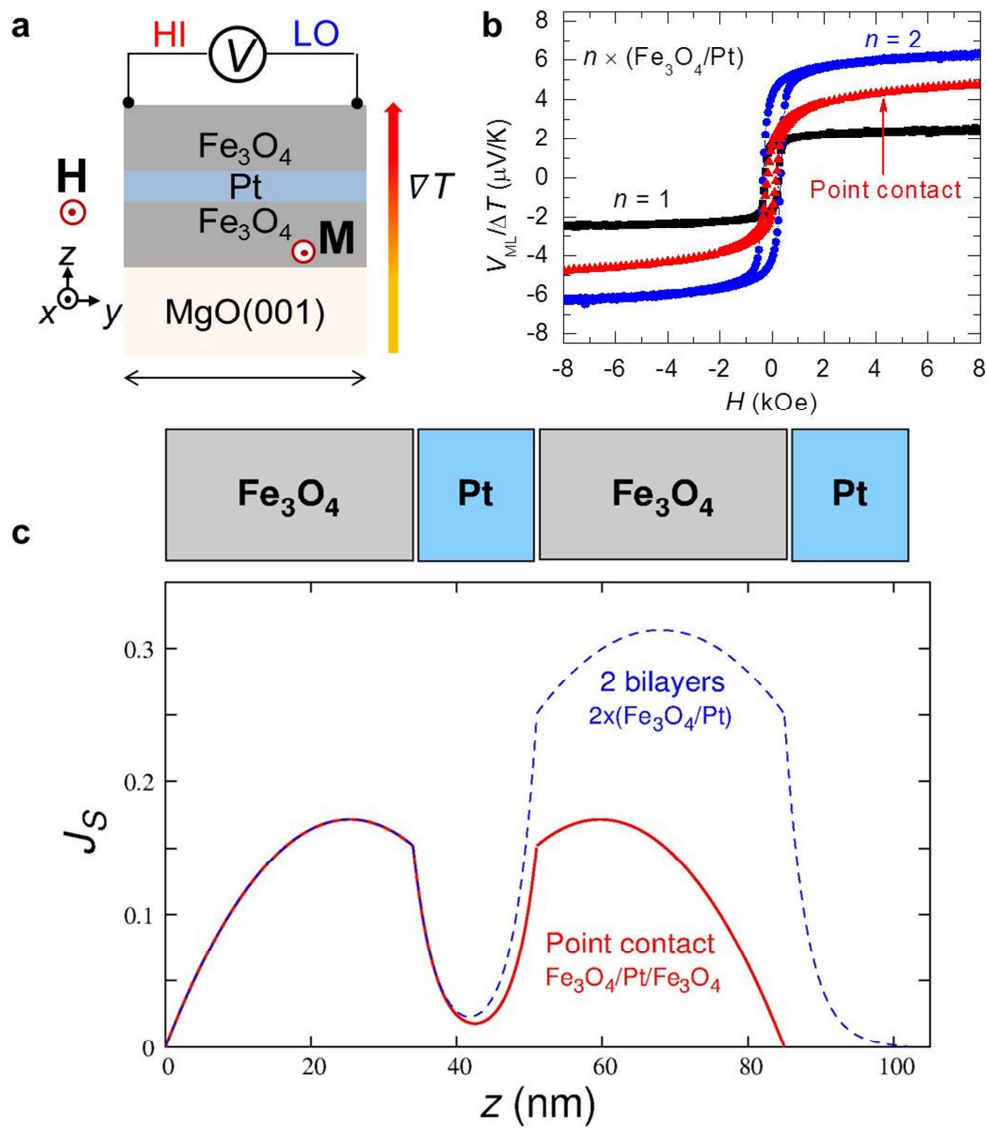

**Supplementary Figure 11 | Spin Seebeck of $Fe_3O_4/Pt/Fe_3O_4$ structure.** Spin Seebeck effect of $Fe_3O_4/Pt/Fe_3O_4$ structure. **(a)** A schematic illustration of the spin Seebeck experiment on the trilayer structure with the point contacts. **(b)** Comparative results of the SSE in $1 \times (Fe_3O_4/Pt)$, $2 \times (Fe_3O_4/Pt)$ and $Fe_3O_4/Pt/Fe_3O_4$ samples. The thicknesses of the layers are $t_{Pt} = 17 \pm 2$ nm and $t_{Fe3O4} = 34 \pm 2$ nm. **(c)** Comparison of the spin current profile for $2 \times (Fe_3O_4/Pt)$ and $Fe_3O_4/Pt/Fe_3O_4$ structures, the spin current strength is normalized by $-L_m \nabla_z T$.

**Supplementary Table 1 | The number of period, thickness and effective thermal conductivity of multilayers.**

Nominal thicknesses: $Fe_3O_4$ (34 nm)/Pt (17 nm)

| Number of period | 1 | 2 | 3 | 4 | 6 |
|---|---|---|---|---|---|
| Total thickness [nm] | 54.6 | 104 | 130 | 196 | 326 |
| Effective thermal conductivity [$Wm^{-1}K^{-1}$] | 2.82 | 2.65 | 3.20 | 2.75 | 2.92 |

Nominal thicknesses: $Fe_3O_4$ (23 nm)/Pt (7 nm)

| Number of period | 1 | 6 | 12 |
|---|---|---|---|
| Total thickness [nm] | 41 | 187 | 341 |
| Effective thermal conductivity [$Wm^{-1}K^{-1}$] | 2.46 | 2.73 | 2.66 |

Nominal thicknesses: 2x[($Fe_3O_4$ (34 nm)/MgO (8 nm)]/[$Fe_3O_4$ (34 nm)/Pt (17 nm)]

| Number of period | 1 |
|---|---|
| Total thickness [nm] | 115 |
| Effective thermal conductivity [$Wm^{-1}K^{-1}$] | 2.63 |

**Supplementary Note 1: Spin Seebeck effect in $Fe_3O_4$/Pt bilayer**

The spin Seebeck effect (SSE) was measured using the longitudinal SSE geometry (Supplementary Fig. 1a), where a thermal gradient ($\nabla T$) is applied in the $z$ direction and the generated spin current is converted to an electric voltage ($V_y$) in the Pt film ($y$ direction) by the inverse spin Hall effect (ISHE), that is measured while a sweeping magnetic field is applied ($x$ direction).

The measurement was initially performed for a single $Fe_3O_4$/Pt bilayer system by measuring the voltage response for different magnitudes of an applied temperature difference ($\Delta T$) (Supplementary Fig. 1b). From the slope of the curve of the voltage at magnetic saturation ($V_y^{sat} = V_y(7\ kOe)$) vs $\Delta T$ (Supplementary Fig. 1c) we can obtain the spin Seebeck coefficient, which is given by the following expression $S_{SSE} = (V_y^{sat}/\Delta T)(L_z/L_y)$.

**Supplementary Note 2: Absence of anomalous Nernst effect in Fe$_3$O$_4$/Pt system**

The SSE is measured by means of the ISHE in the Pt layer, which generates an electric field ($\mathbf{E}_{ISHE}$), that is expressed according to the relation

$$\mathbf{E}_{ISHE} = (\theta_{SH}\rho)\,\mathbf{J}_s \times \boldsymbol{\sigma} \qquad (S1)$$

where $\theta_{SH}$, $\rho$, $\mathbf{J}_s$, $\boldsymbol{\sigma}$ denote the spin-Hall angle and electric resistivity of the Pt layer, spin current and spin-polarization (parallel to the magnetization, **M**) respectively. In the longitudinal SSE measurement geometry the generated spin current is parallel to the applied thermal gradient $\nabla T$. This experimental configuration has the same geometry as the one used for the observation of the anomalous Nernst effect (ANE)

$$\mathbf{E}_{ANE} \propto \nabla T \times \mathbf{M} \qquad (S2)$$

Therefore, care must be taken when measuring materials where the ANE effect is also present. To estimate the suppression of the anomalous Nernst effect (ANE) of Fe$_3$O$_4$ upon placement of a Pt layer, we will consider the expression previously derived from electron transport theory (*25*):

$$E_y = \frac{r}{1+r} E_{ANE} \qquad (S3)$$

which gives the value of the electric field in the *y* direction as a function of the parameter $r = \dfrac{\rho_{Pt}}{\rho_{Fe_3O_4}} \dfrac{t_{Fe_3O_4}}{t_{Pt}}$ and the anomalous Nernst field ($E_{ANE}$) measured in the Fe$_3$O$_4$ film. Considering a resistivity value of $\rho_{Pt} = 1.39 \times 10^{-7}$ $\Omega$m ($3.77 \times 10^{-7}$ $\Omega$m) for a Pt film with $t_{Pt} = 17$ nm (7 nm) and $\rho_{Fe3O4} = 6.96 \times 10^{-5}$ $\Omega$m for a 34 nm thick Fe$_3$O$_4$ film, we obtain that the ANE signal in the

Fe$_3$O$_4$/Pt($t_{Pt}$) bilayer with $t_{Pt}$ = 17 nm (7 nm) is reduced to 0.4 % (3 %) of the ANE signal in the Fe$_3$O$_4$ layer. This gives a contribution of about 0.4 % (1 %) to the total observed signal $V_y = E_y L_y$. This is a strong indication that the SSE is the dominant contribution to the observed voltage.

Furthermore, in order to experimentally check the absence of the ANE due to magnetic proximity in the Pt layer. Measurements were performed in the in-plane magnetized, IM (Supplementary Fig. 2a) and perpendicular magnetized, PM (Supplementary Fig. 2b) configurations (*20, 26*), these are performed by interchanging the directions of the applied thermal gradients $\nabla T$ and magnetic field **H**, while keeping all the parameters: thermal gradient, magnetic field and direction of measured voltage orthogonal with one another. Supplementary Fig. 2c shows the results obtained for a Fe$_3$O$_4$/Pt bilayer measured in the IM configuration, also shown is an estimation of the contribution of the ANE to the total SSE measured signal, as calculated above. When the sample is measured in the PM configuration, the observation of SSE is forbidden by measurement geometry (Supplementary Fig. 2d), since the direction of the spin pumping across the Fe$_3$O$_4$/Pt interface is parallel to the direction of magnetization (*z* direction), see equation (S1). This result demonstrates that the signal measured in the Fe$_3$O$_4$/Pt bilayer is highly dominated by the SSE.

**Supplementary Note 3: Spin Seebeck effect of thick magnetite layer**

In order to rule out the dependence of the SSE on the thickness of the ferromagnetic material, as a possible source of the giant SSE observed in the multilayer system, we performed measurements on a $Fe_3O_4$(278 nm)/Pt(14 nm) bilayer, where the $Fe_3O_4$ thickness is larger than the total thickness of ferromagnetic material in any of the multilayer samples with $n = 6$ bilayers.

Supplementary Fig. 3 shows the obtained results, it can be clearly observed that the SSE for the single bilayer with a $Fe_3O_4$ thickness of 278 nm is much lower than the results obtained for any of the samples with $n = 6$. This points out to the importance of the increased number of interfaces in the enhancement of the SSE observed in multilayers.

**Supplementary Note 4: Measurement technique to evaluate thermal conductivity of layered structures.**

The thermal conductivity of $Fe_3O_4$/Pt multilayered structures in the cross-plane direction was measured by using the time domain thermoreflectance (TDTR) method[1-4]. All the measurements were carried out at room temperature. The TDTR method is well established and has been successfully applied to a number of thin layered materials to measure the thermal conductivity of the thin layers and the interfacial thermal conductance[5,6]. For the TDTR measurements, an Al thin film was deposited on the sample surface as a transducer. Therefore, the sample consists of 3 parts: MgO substrate, $Fe_3O_4$/Pt multilayer, and Al layer. We have also prepared a reference sample by simultaneously depositing Al directly onto the fused quartz substrate to measure the Al layer thickness using TDTR.

A Ti:sapphire laser with pulse duration of 140 fs and a repetition frequency of 80 MHz was used to generate repetitive pulse heating on the transducer surface and also to measure its temperature. The laser beam is split into pump (400 nm, 100 mW, and 50 μm $1/e^2$-radius) and probe (800 nm, 10 mW, and 5 μm $1/e^2$-radius) beams using polarization beam splitter and nonlinear optical crystal, and the transient thermal response to the heat pulse was resolved in time domain by an optical *pump-probe* scheme. The variations in reflectivity of the sample surface measured using the probe beam can be translated into temperature information through thermoreflectance. Detection of the small reflectance variations is realized by modulating the pump beam at 11 MHz by using an electro-optic modulator and detecting the corresponding fundamental frequency component of the probe signal using a lock-in-amplifier. Supplementary Fig. 4 shows the schematic diagram of the TDTR setup.

The ratio of the in-phase *(X)* and out-of phase *(Y)* signals from the lock-in-amplifier was fitted to an analytical heat conduction model to obtain thermal conductivity of the multilayers. The method of extracting the unknown thermal properties (thermal conductivity and interfacial thermal conductance) of multilayers from detected TDTR signals using the heat conduction model is well established and has been described in detail in the literature[2-4] (we closely followed the approach of Supplementary Ref. 4). In this work, the multilayer is modelled as one medium with an effective thermal conductivity.

The Al film thickness was obtained by performing the TDTR measurements of the reference sample with the thickness as the fitting parameter[7]. The thickness was also checked by the acoustic echo signal[8]. The thermal conductivity of the Al film was obtained by the Wiedemann-Franz law[7]. The volumetric heat capacity of the multilayer part was obtained by measuring the thickness of each Pt and $Fe_3O_4$ layer by the transmission electron microscope (TEM). For the heat capacity and thermal conductivity of the MgO substrate, we took the bulk values from literature. These procedures reduce the unknown parameters to effective thermal conductivity of multilayer part and interfacial thermal conductance at Al/multilayer (Al/Pt) interface. Note that the thermal resistance of the MgO/Fe$_3$O$_4$ interface is included in the multilayer thermal resistance when extracting the effective thermal conductivity of the multilayer. It was confirmed that this two-parameter fitting gives rise to sufficiently selective and large parameter sensitivity. The fitting was performed for the TDTR signal at times later than 100 ps from the time of the impulse, to neglect the effect of electron-phonon coupling and to ensure that the metal transducer film had reached local equilibrium.

Thermal conductivities of 10 samples listed in Supplementary Table 1 were measured by TDTR. Supplementary Fig. 5 shows the fitting results of Fe$_3$O$_4$ (34 nm)/Pt (17 nm) multilayers with the number of Fe$_3$O$_4$/Pt period $n$ = 1, 2, and 4. Supplementary Figs. 6a and 6b summarize the effective thermal conductivity for Fe$_3$O$_4$/Pt multilayer as a function of $n$ for the cases of Fe$_3$O$_4$ (34 nm)/Pt (17 nm) and Fe$_3$O$_4$ (23 nm)/Pt (7 nm), respectively. The results show that the dependence of the effective thermal conductivity of the multilayer on the number of period is small and thus is not the reason for the strong enhancement of the SSE.

**Supplementary Note 5: Theory of the longitudinal SSE in multilayer systems**

In this section, we outline our theoretical analysis of the SSE in a multilayer system composed of a ferromagnet (F, in the experiments $Fe_3O_4$) and non-magnetic metal (N, in the experiments Pt). Here, two types of spin currents play key roles in the multilayer SSE; conduction-electron spin current in the N layer and magnon spin current in the F layer. We consider a multilayer system composed of $n$ sets of a F/N unit, as shown in Supplementary Fig. 7. To describe the conduction-electron spin current in the N layer, we start from the Bloch-Torrey equation with spin diffusion[9], and project the spin accumulation $\delta s$ onto the $x$-axis (∥ external magnetic field). Following the procedure given in Supplementary Ref. 10, we obtain the spin diffusion equation[11]:

$$\nabla^2 \delta s = \frac{1}{\lambda^2} \delta s, \quad (S4)$$

$$\mathbf{J}_s = -D_s \nabla \delta s, \quad (S5)$$

where $\lambda$ is the spin diffusion length, $\mathbf{J}_s$ is the conduction-electron spin current, and $D_s$ is the spin diffusion coefficient.

To describe the magnon current in the F layer, we begin with the Landau-Lifshitz-Gilbert equation, next introduce the spin-wave approximation, and then project the magnon current onto the spin quantization axis (i.e., $x$-axis) as $j_m = \hat{x} \cdot D_{ex} \hat{\mathbf{M}} \times \nabla \hat{\mathbf{M}}$, where $\hat{\mathbf{M}}$ is the unit vector of the magnetization direction and $D_{ex}$ is the exchange stiffness. Following the procedure given in Supplementary Ref. 12, we introduce the magnon accumulation $\delta m$ and obtain the magnon diffusion equation [see equation (68) in Supplementary Ref. 12]:

$$\nabla^2 \delta m = \frac{1}{\Lambda^2} \delta m, \tag{S6}$$

$$\mathbf{J}_m = -L_m \nabla T - D_m \nabla \delta m, \tag{S7}$$

where $\Lambda$ is the magnon diffusion length, $\mathbf{J}_m = <\mathbf{j}_m>$ is the thermal average of the magnon spin current, $D_m (L_m)$ is the magnon diffusion (magnon thermal diffusion) coefficient. Note that the same equations can be obtained by using Boltzmann approach to the magnon dynamics[13].

In order to determine the spin current profile in the multilayer system, we need to apply physical boundary conditions satisfied at the F/N interface. To simplify the argument, we assume that the spin accumulation and magnon accumulation depend only on the coordinate $z$ whose direction is perpendicular to the interface (see Supplementary Fig. 7). A temperature gradient is applied perpendicular to the interface, such that both the spin current and magnon current are parallel to the $z$ direction, i.e., $\mathbf{J}_s = J_s \hat{z}$ and $\mathbf{J}_m = J_m \hat{z}$. If we take the $+z$-axis as the positive direction of currents, we can write down the following boundary conditions for $J_s$ and $J_m$ at a F/N interface:

$$\begin{pmatrix} J_s \\ J_m \end{pmatrix} = \begin{pmatrix} \sigma_s, & -\varsigma_s \\ \varsigma_m, & -\sigma_m \end{pmatrix} \begin{pmatrix} \delta m \\ \delta s \end{pmatrix}, \tag{S8}$$

where we introduced four transport coefficients $\sigma_s$, $\sigma_m$, $\varsigma_s$, and $\varsigma_m$. Note that equation (S8) represents a generalization of the boundary conditions previously used in the literature. Takahashi et al.[14] used a condition $\varsigma_s = \varsigma_m = 0$. Zhang et al.[13] used a condition $\sigma_s = \varsigma_m$ and $\sigma_m = \varsigma_s$, so that we have the continuity of spin current ($J_s = J_m$) at the interface. Essentially the same condition is used by Hoffman et al.[15] as well.

We apply the boundary conditions (S8) to the multilayer system under consideration (Supplementary Fig. 7), where each F (N) layer has a thickness $t_F$ ($t_N$). We introduce coordinates $Z_i$ and $z_i$ to specify a position within a layer, so that any position in the system can be written as

$$z = \begin{cases} (t_F + t_N)(i-1) + Z_i & \text{for } z \in F(i) \\ (t_F + t_N)(i-1) + t_F + z_i & \text{for } z \in N(i) \end{cases} \quad (S9)$$

by choosing a proper value of $i$. The magnon accumulation and the spin accumulation in the $i$-th unit $F(i)/N(i)$ can be written as

$$\delta m^{(i)}(z) = \langle \delta M^{(i)} \rangle \cosh\left(\frac{Z_i}{\Lambda}\right) + A^{(i)} \sinh\left(\frac{Z_i}{\Lambda}\right),$$

$$\delta s^{(i)}(z) = \langle \delta S^{(i)} \rangle \cosh\left(\frac{z_i}{\lambda}\right) + B^{(i)} \sinh\left(\frac{z_i}{\lambda}\right), \quad (S10)$$

where, by using the boundary conditions at the $F(i)/N(i)$ interface, the two coefficients $A^{(i)}$ and $B^{(i)}$ can be represented in terms of $\langle \delta M^{(i)} \rangle$ and $\langle \delta S^{(i)} \rangle$ as

$$A^{(i)}[\langle \delta M^{(i)} \rangle, \langle \delta S^{(i)} \rangle] = \frac{1 - [\sinh(t_F/\Lambda) + F_m \cosh(t_F/\Lambda)]\langle \delta M^{(i)} \rangle + G_m \langle \delta S^{(i)} \rangle}{\cosh(t_F/\Lambda) + F_m \sinh(t_F/\Lambda)},$$

$$B^{(i)}[\langle \delta M^{(i)} \rangle, \langle \delta S^{(i)} \rangle] = -G_m [\sinh(t_F/\Lambda) A^{(i)} + F_m \cosh(t_F/\Lambda)\langle \delta M^{(i)} \rangle] + F_s \langle \delta S^{(i)} \rangle. \quad (S11)$$

Thus, if the two coefficients $\langle \delta M^{(i)} \rangle$ and $\langle \delta S^{(i)} \rangle$ are given, the magnon and spin currents are determined by

$$J_m^{(i)}(z) = -L_m \nabla_z T - \frac{D_m}{\Lambda}\left[\langle \delta M^{(i)} \rangle \sinh\left(\frac{Z_i}{\Lambda}\right) + A^{(i)} \cosh\left(\frac{Z_i}{\Lambda}\right)\right],$$

$$J_s^{(i)}(z) = -\frac{D_s}{\lambda}\left[\langle \delta S^{(i)} \rangle \sinh\left(\frac{z_i}{\lambda}\right) + B^{(i)} \cosh\left(\frac{z_i}{\lambda}\right)\right]. \quad (S12)$$

We can derive a recursion relation between $(\langle\delta M^{(i)}\rangle, \langle\delta S^{(i)}\rangle)$ and $(\langle\delta M^{(i+1)}\rangle, \langle\delta S^{(i+1)}\rangle)$ using the boundary conditions at the interface between a F($i$)/N($i$) unit and a F($i+1$)/N($i+1$) unit:

$$\begin{pmatrix} a_{11}, & a_{12} \\ 0, & a_{22} \end{pmatrix} \begin{pmatrix} \langle\delta M^{(i)}\rangle \\ \langle\delta S^{(i)}\rangle \end{pmatrix} = \begin{pmatrix} b_{11}, & 0 \\ b_{12}, & b_{22} \end{pmatrix} \begin{pmatrix} \langle\delta M^{(i+1)}\rangle \\ \langle\delta S^{(i+1)}\rangle \end{pmatrix} + \begin{pmatrix} c_1 \\ c_2 \end{pmatrix}, \quad (S13)$$

where each matrix element is given by

$$\begin{aligned}
a_{11} &= b_{22} = \frac{-1}{\cosh(t_F/\Lambda) + F_m \sinh(t_F/\Lambda)}, \\
a_{22} &= b_{11} = \frac{1}{\cosh(t_N/\lambda) + F_s \sinh(t_N/\lambda)}, \\
a_{12} &= \frac{1}{G_s}\frac{\tanh(t_N/\lambda) + F_s}{1 + F_s \tanh(t_N/\lambda)} + \frac{F_s}{G_s} - \frac{G_m \tanh(t_F/\Lambda)}{1 + F_m \tanh(t_F/\Lambda)}, \\
b_{21} &= \frac{1}{G_m}\frac{\tanh(t_F/\Lambda) + F_m}{1 + F_m \tanh(t_F/\Lambda)} + \frac{F_m}{G_m} - \frac{G_s \tanh(t_N/\lambda)}{1 + F_s \tanh(t_N/\lambda)}, \\
c_1 &= \frac{\tanh(t_F/\Lambda)}{1 + F_m \tanh(t_F/\Lambda)}, \\
c_2 &= \frac{1}{G_m}\left(1 - \frac{1}{\cosh(t_F/\Lambda) + F_m \sinh(t_F/\Lambda)}\right).
\end{aligned} \quad (S14)$$

Here, $G_s = \sigma_s \lambda/D_s$, $G_m = \sigma_m \Lambda/D_m$, $F_s = \varsigma_s \lambda/D_s$, $F_m = \varsigma_m \lambda/D_m$, and the coefficients $\langle\delta M^{(i)}\rangle$ and $\langle\delta S^{(i)}\rangle$ are normalized by $-\Lambda L_m \nabla_z T/D_m$.

Now we demonstrate that a multilayer structure enhances the SSE. To do so, we solve the recursion relation [equation (S13)] numerically, together with the boundary condition that the spin current (magnon current) at the topmost surface (the bottommost surface) vanishes, i.e., $J_s|_{\text{topmost}} = 0$ $(J_m|_{\text{bottommost}} = 0)$. We use the continuity of the spin/magnon current across the interface [S8, S10] to simplify the argument. Once $\langle\delta M^{(i)}\rangle$ and $\langle\delta S^{(i)}\rangle$ are obtained, other coefficients $A^{(i)}$ and $B^{(i)}$ can be determined using equation (S11), and hence we can calculate the spin current profile.

In Supplementary Fig. 8a, we plot the spin current profile for a F(30nm)/N(5nm) multilayer system, where the lateral dimension is assumed to be of the order of several millimeters, and the spin/magnon current strength is normalized by $-L_m \nabla_z T$. To highlight the multilayer effect, we here pick up a situation where the magnon diffusion length (spin diffusion length) is twice as large as the thickness of the F layer (N layer), i.e., $\lambda = 10$ nm and $\Lambda = 60$ nm, and set $G_s = G_m = F_s = F_m = 5$. Note that, the spin current in N at the toplayer inevitably vanishes at the topmost surface, whereas the spin current in N at the interlayer remains a nonzero value throughout the multilayer, resulting in an enhancement of the spin current in N.

When discussing the SSE in such a multilayer system where the lateral dimension is about six orders of magnitude larger than the thickness of the F/N unit, a care is necessary to interpret the signal, because the out-of-plane electric resistance of F is negligibly small in comparison with the in-plane resistance. In this situation, even if the F layer is made of an insulating ferromagnet, simply by measuring the voltage between two point contacts to N at the topmost layer, we detect the voltage signal that is produced in N at the interlayer. Such a voltage signal $V_{\mathrm{ISHE}}$ is given, via the inverse spin Hall effect, as

$$V_{\mathrm{ISHE}} \propto \theta_{\mathrm{SH}} \langle J_s \rangle \rho_{\mathrm{N}}, \tag{S15}$$

where $\theta_{\mathrm{SH}}$ is the spin-Hall angle, $\rho_{\mathrm{N}}$ is the resistivity of the N layer, and

$$\langle J_s \rangle = \frac{1}{t_{\mathrm{N}} n} \sum_{i=1}^{n} \int_{z_i=0}^{t_{\mathrm{N}}} dz J_s^{(i)}(z) \tag{S16}$$

is the spin current averaged over all N layers.

Supplementary Fig. 8b shows the averaged spin current $\langle J_s \rangle$ as a function of the number of bilayers, $n$. Clearly, one can see that the SSE is amplified by increasing $n$. The physics behind is that, thanks to the multilayer structure, spin current in N at the interlayer acquires new length scale and boundary value, thereby resulting in an enhancement of the spin current. The calculation thus demonstrates that a multilayer structure can enhance the SSE.

Keeping in mind the spin current profile shown in Supplementary Fig. 8a, let us next discuss the enhancement of the SSE observed in the $Fe_3O_4$(23 nm)/Pt(7 nm) multilayer system as well as in the $Fe_3O_4$(34 nm)/Pt(17 nm) multilayer system. To discuss the experimental result, we first fix the spin diffusion length $\lambda$ in Pt and the magnon diffusion length $\Lambda$ in $Fe_3O_4$. Following Supplementary Ref. 16 we adopt a value $\lambda = 3$ nm for Pt, and concerning $\Lambda$ we use $\Lambda = 40$ nm which gives a reasonable fit to the dependence on the thickness of $Fe_3O_4$ layer of the SSE in a single $Fe_3O_4$/Pt bilayer.

In Supplementary Fig. 9, we plot the averaged spin current as a function of the number of $Fe_3O_4$/Pt bilayers. The calculation reproduces the main features of the multilayer effects on the spin Seebeck signal, i.e., the observed SSE is amplified upon increasing the number of $Fe_3O_4$/Pt bilayers. The calculation shows that the averaged spin current in the $Fe_3O_4$(23 nm)/Pt(7 nm) multilayer sample is about 1.5 times larger than that in $Fe_3O_4$(34 nm)/Pt(17 nm). If we take into account the fact that the resistivity of the Pt layer in the $Fe_3O_4$(23 nm)/Pt(7 nm) multilayer sample can be twice as large as that in the $Fe_3O_4$(34 nm)/Pt(17 nm) multilayer sample (see Fig. 3 in Supplementary Ref. 16), we can explain the experimental observation, based on equation (S15) that the SSE in the $Fe_3O_4$(23 nm)/Pt(7 nm) multilayer sample is three times larger than that in the $Fe_3O_4$(34 nm)/Pt(17 nm) multilayer sample.

**Supplementary Note 6: Resistance vs number of Fe$_3$O$_4$/Pt bilayers (*n*).**

The resistance measurements were performed using a 4 probe geometry, with the electrical contacts placed on the topmost Pt layer. As a consequence of the sample geometry in which the lateral dimensions of the sample ($L_x$, $L_y$ ~ mm) are 6 orders of magnitude larger than the dimensions of the films (nm). In this situation, even if magnetite is two orders of magnitude more resistive than Pt, the out-of-plane resistance is negligible small compared to the in-plane resistance, thus resulting in a parallel contact of all the Pt layers (Supplementary Fig. 10) and the resistance is reduced as the number of bilayers (*n*) increases.

**Supplementary Note 7: Spin Seebeck measurements on Fe$_3$O$_4$/Pt/Fe$_3$O$_4$ trilayers.**

The SSE measurements were performed on structures in which the top Pt layer is removed, resulting in a trilayer heterostructure formed by a Pt interlayer sandwiched between two Fe$_3$O$_4$ layers (Supplementary Fig. 11a). Since the thickness of the Fe$_3$O$_4$ layers is about $10^6$ times shorter than the in-plane length, the out-of-plane electric resistance (*z* direction) of Fe$_3$O$_4$ is negligibly small in comparison with the in-plane resistance (*y* direction). Therefore, we can detect the spin Seebeck voltage generated in the Pt interlayer by simply measuring the voltage between the two point contacts placed on the topmost layer.

Supplementary Fig. 11b shows the comparative results obtained for 1 × (Fe$_3$O$_4$/Pt), 2 × (Fe$_3$O$_4$/Pt) and the point contact sample (Fe$_3$O$_4$/Pt/Fe$_3$O$_4$). The SSE in the point contact sample shows an intermediate value between the ones obtained for *n* = 1 and 2. This is consistent with an estimation of the spin current profile across the structures (Supplementary Fig. 11c) based on our theoretical model presented above and in agreement with a model presented in Ref. 15, where it is shown that the spin Seebeck coefficient of a N1/F/N2 structure is increased when N1 and N2 are good spin sinks (i. e. Pt) in comparison to the value estimated when N1 is a poor spin sink.

This result further proves that we are measuring an average spin Seebeck contribution from all the Pt layers of the multilayer structures even if the electrical contacts are placed on the topmost Pt layer only.